\documentclass[a4paper,11pt]{article}
\usepackage{pos}
\usepackage{subfig} 

\title{Higgs boson pair production at N$^3$LO QCD}

\author[a]{Long-Bin Chen}
\author[b,c]{Hai Tao Li}
\author*[d]{Hua-Sheng Shao}
\author[e]{Jian Wang}

\affiliation[a]{School of Physics and Electronic Engineering, Guangzhou University, Guangzhou 510006, China}

\affiliation[b]{HEP Division, Argonne National Laboratory, Argonne, Illinois 60439, USA}

\affiliation[c]{Department of Physics \& Astronomy, Northwestern University, Evanston, Illinois 60208, USA }

\affiliation[d]{Laboratoire de Physique Th\'eorique et Hautes Energies (LPTHE), UMR 7589, Sorbonne Universit\'e et CNRS,
4 place Jussieu, 75252 Paris Cedex 05, France}

\affiliation[e]{School of Physics, Shandong University, Jinan, Shandong 250100, China}

\emailAdd{chenlb@gzhu.edu.cn}
\emailAdd{haitao.li@northwestern.edu}
\emailAdd{huasheng.shao@lpthe.jussieu.fr}
\emailAdd{j.wang@sdu.edu.cn}

\abstract{Understanding the Higgs potential by measuring its self-interactions is fundamental in answering several big questions, such as electroweak symmetry breaking, electroweak baryogenesis, electroweak phase transition, and electroweak vacuum stability. The most promising way to probe the Higgs potential is to detect Higgs boson pair final state at high-energy colliders. In this talk, we report a recent perturbative calculation for the di-Higgs gluon-fusion process by taking into account N$^3$LO QCD radiative corrections in the approximation of infinite top quark mass limit. Finite top quark mass effects are also incorporated with several approximate schemes, which are known to be crucial in phenomenological applications. We show a very good asymptotic perturbative convergence at $\mathcal{O}(\alpha_s^5)$, and demonstrate that the remaining scale uncertainty is only at percent level.}

\FullConference{%
  40th International Conference on High Energy physics - ICHEP2020\\
  July 28 - August 6, 2020\\
  Prague, Czech Republic (virtual meeting)
}

\begin{document}
\maketitle

\vspace{-0.5cm}
\section{Introduction}
\vspace{-0.3cm}

\begin{figure}[!ht]
\vspace{-0.5cm}
  \centering
  \subfloat[First-order phase transition]{\includegraphics[width=0.30\textwidth,draft=false]{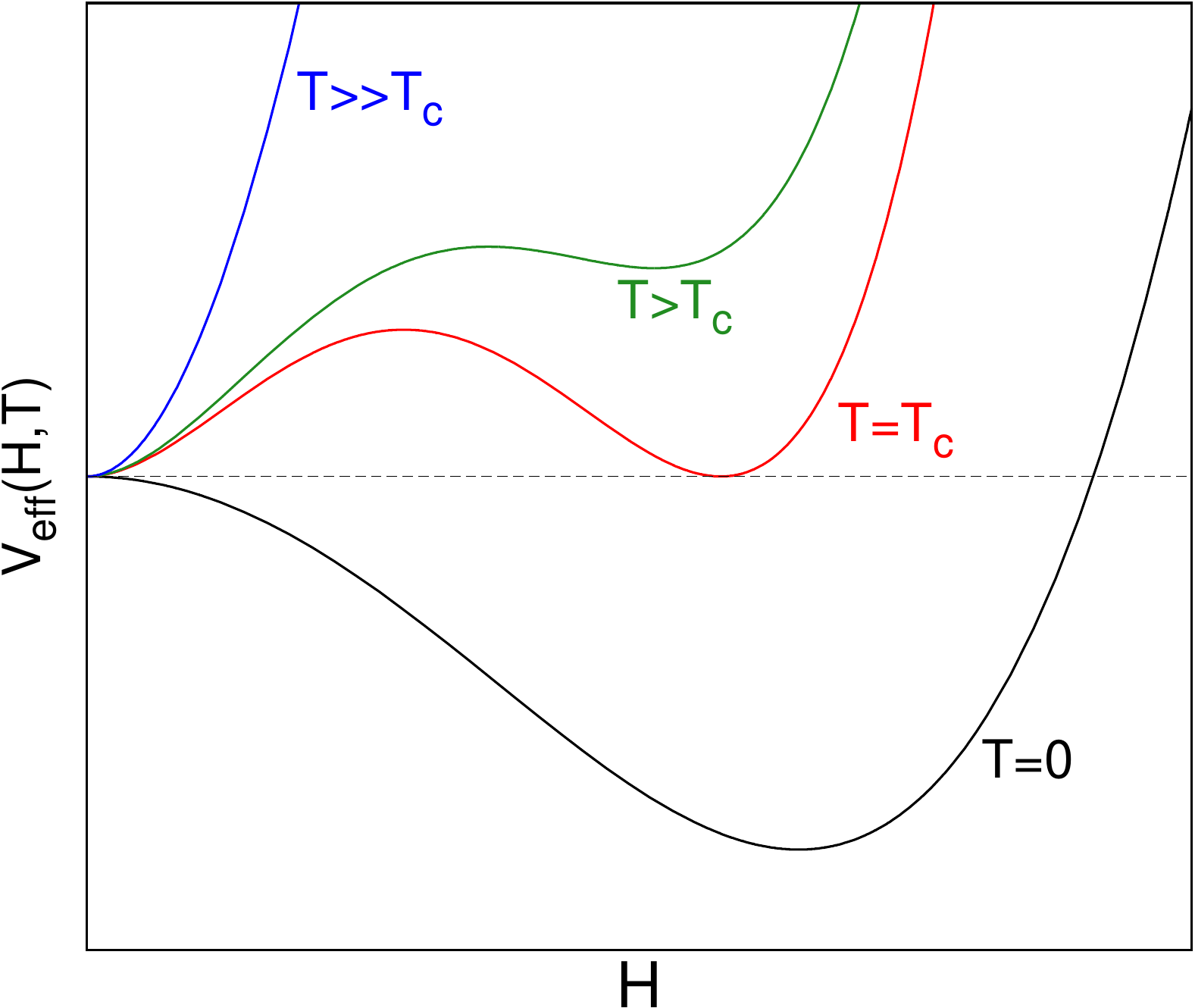}\label{fig:FOPS}}
  \subfloat[Second-order phase transition]{\includegraphics[width=0.30\textwidth,draft=false]{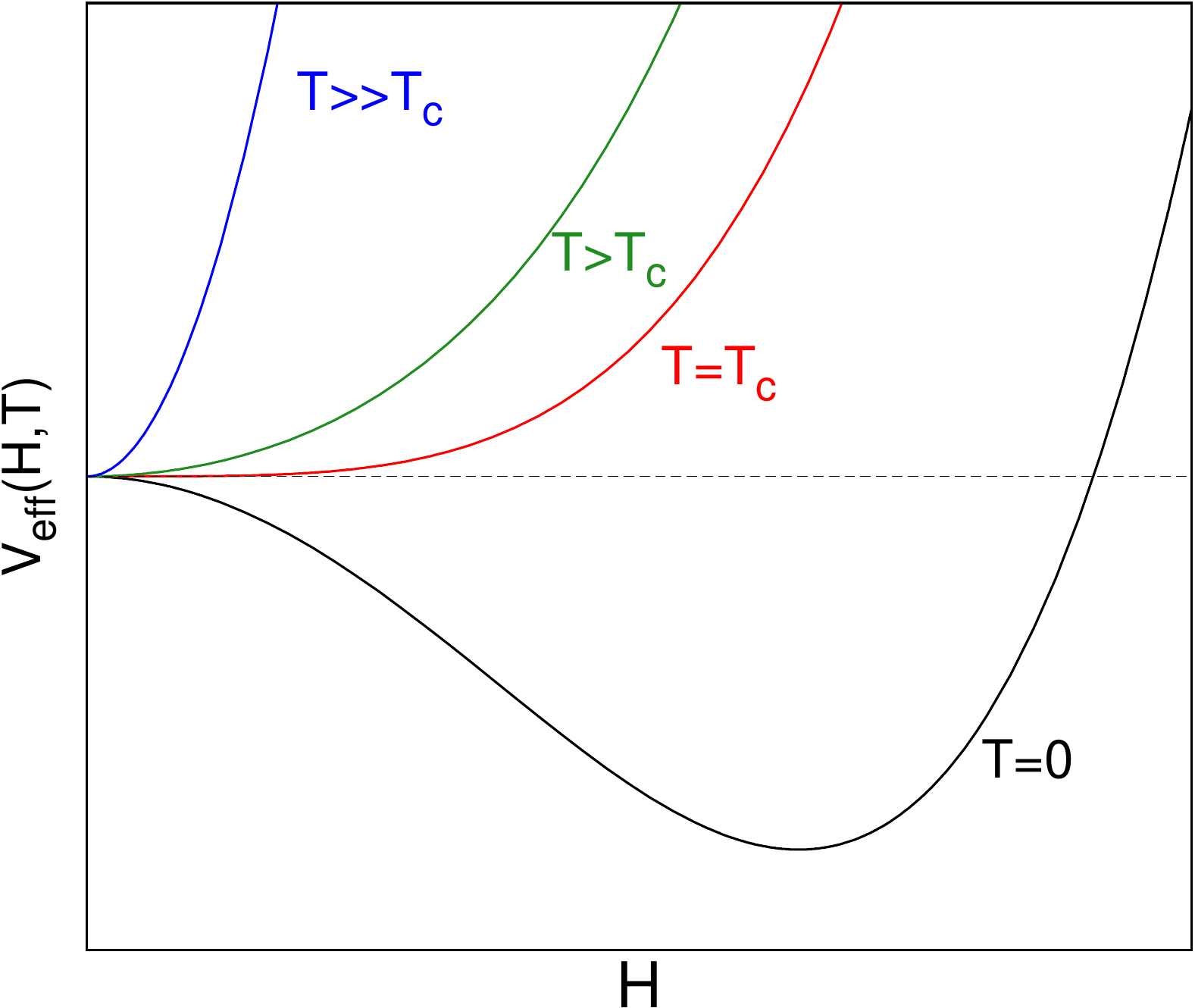}\label{fig:SOPS}}
  \subfloat[Vacuum (in)stability]{\includegraphics[width=0.30\textwidth,draft=false]{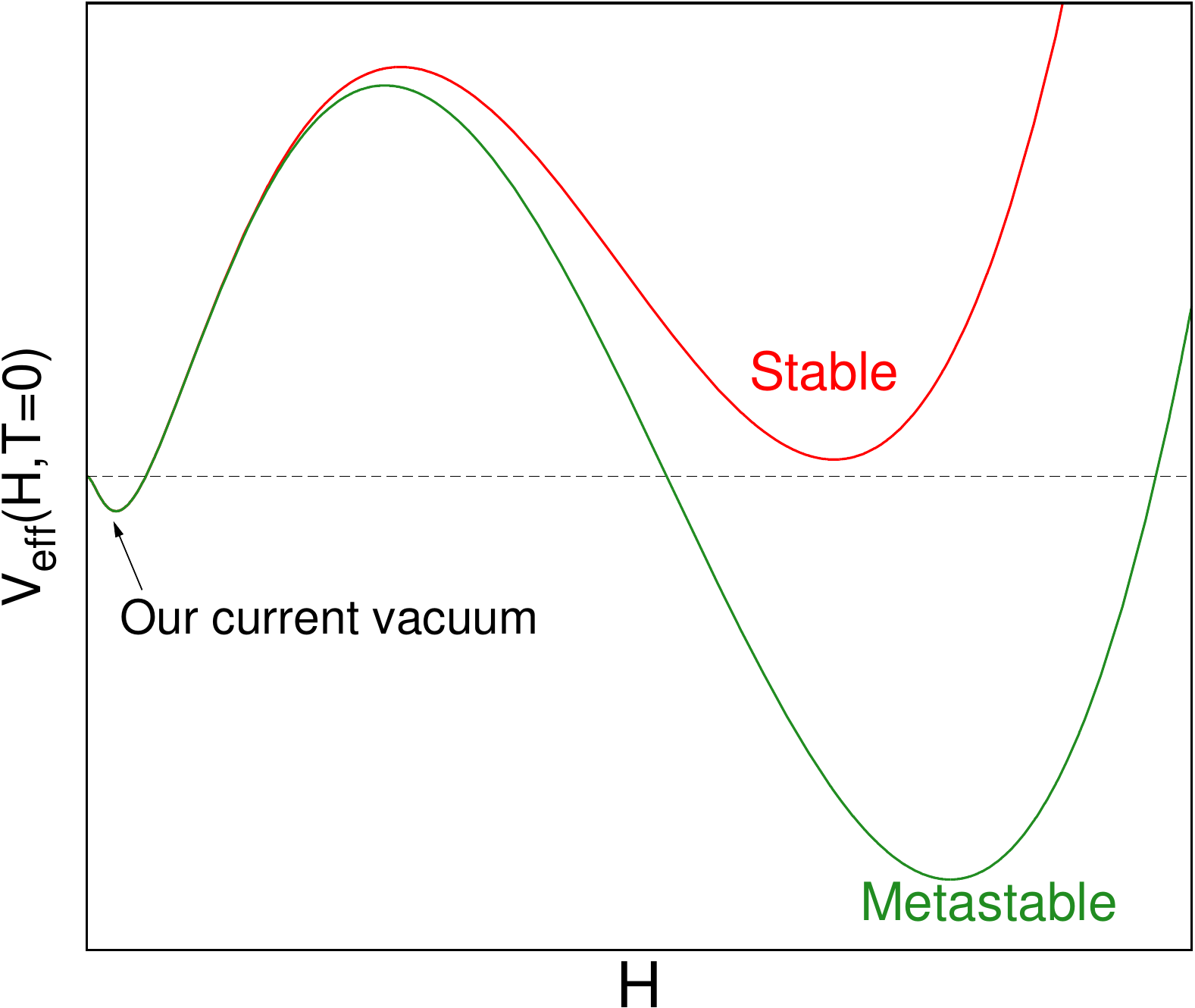}\label{fig:VS}}
  \vspace{-0.3cm}
  \caption{Schematic illustrations of Higgs potential at finite temperature for (a) first-order phase transition and (b) second-order phase transition, and of (c) Higgs potential at zero temperature.
  \label{fig:plot}}
  \vspace{-0.4cm}
\end{figure}

It is widely acknowledged that the discovery of the Higgs boson at the LHC has singled out a unique route to deepen our understanding of the fundamental laws of nature at subatomic scales. One of the most peculiar features in the Standard Model (SM) is the shape of the Higgs potential, which drives the electroweak symmetry breaking and therefore the mass acquirements of the observed particles. Many people believe that there must be a (unknown) microscopic interpretation of the Higgs potential, an analogy for the Bardeen-Cooper-Schrieffer theory to the superconductivity. The Higgs potential is also crucial in answering several other big questions related to cosmology.  In the thermal history of early universe, different Higgs potentials predict that our universe has undergone different types of electroweak phase transitions. Lattice calculations show that there is no phase transition (i.e., a crossover) in the setup of SM~\cite{Kajantie:1996mn,Csikor:1998eu}. Generic Higgs sectors in many extensions of the SM (BSM) predict phase transitions in the electroweak epoch, in which most of them are first order (like Fig.~\ref{fig:FOPS}) while the rest could be second order (like Fig.~\ref{fig:SOPS}, known with infinite correlation length). The first-order phase transition is very interesting to explain the matter-antimatter asymmetry in the universe (known as {\it baryogenesis})~\cite{Shaposhnikov:1987tw}. The first-order phase transition may also leave detectable cosmic imprints, such as primordial gravitational waves~\cite{Witten:1984rs,Hogan:1986qda} and magnetic fields~\cite{Grasso:2000wj}. The Higgs potential also determines the stability of our current vacuum as illustrated in Fig.~\ref{fig:VS}. If it is like the green curve in the figure, as what most likely the case in the SM~\cite{Degrassi:2012ry}, our universe may decay to a lower vacuum through quantum tunneling at some point.

Among all possible Higgs self couplings that characterise the Higgs potential, the Higgs trilinear coupling $\lambda_{hhh}$ is understood as the most viable quantity that can be measured precisely. The precision of $\lambda_{hhh}$ is therefore crucial to disentangle the different phase transitions, and to probe BSM dynamics that might be at very high scales~\cite{Chang:2019vez}. The determinations of $\lambda_{hhh}$, either \textit{indirectly} through single Higgs or \textit{directly} via di-Higgs production, at future experiments are expected to be at the level of $50\%$ ($20\%,5\%$) accuracy at HL-LHC (future $e^+e^-$ machines, FCC-hh) with respect to its SM value~\cite{deBlas:2019rxi}. In particular, it has been suggested that a 100 TeV pp collider could be an ultimate precision machine for this quantity~\cite{Mangano:2020sao}. 

In this talk, we will focus on the Higgs boson pair process at pp colliders, such as LHC, HE-LHC and FCC-hh, which is dominantly produced via gluon-gluon fusion (ggF). Our discussion will also be mainly in the context of the SM. The precision theoretical calculations of $gg\to hh$ in the literature mainly follow two different routes. The first one is to carry out computations with full top-quark mass $m_t$ dependence, which are very hard to improve because the lowest order is already a loop-induced process. Thus, the full next-to-leading order (NLO) QCD corrections, as well as their matching to parton shower (PS) programmes, were only known recently~\cite{Borowka:2016ehy,Borowka:2016ypz,Baglio:2018lrj,Davies:2019dfy,Heinrich:2017kxx,Jones:2017giv,Heinrich:2019bkc,Baglio:2020ini}. Several lessons we have learned from these studies. The intrinsic theoretical uncertainty only from renormalisation and factorisation scale variations may be underestimated significantly due to the ignorance of other possible sources, such as those stemming from the ambiguities in the top quark mass scheme, the PS matching scheme, and the PS shower scale choices. Besides, the usual reasonable approximations to extend top-quark mass $m_t$ corrections by rescaling full $m_t$ Born and even real radiation~\cite{Maltoni:2014eza} can fail the true QCD corrections significantly. The second approach usually being taken is to approach the infinite/heavy/large top quark mass limit ($m_t\to \infty$), which is well motivated in the single Higgs case. The obvious advantage of this approach is that the process starts to feature non-zero tree-level contribution, whose precision is easier to be improved. Indeed, the NLO computation has been known since two decades~\cite{Dawson:1998py}. Next-to-next-to-leading order (NNLO) was also available~\cite{deFlorian:2013uza,deFlorian:2013jea,Grigo:2014jma,deFlorian:2016uhr} for a while, and recently we have carried out the first next-to-next-to-next-to-leading order (N$^3$LO) calculation~\cite{Chen:2019lzz,Chen:2019fhs}, which is the main content of this talk. In the same approximation, the soft-gluon resummation effects have also been considered in Refs.~\cite{Shao:2013bz,deFlorian:2015moa,deFlorian:2018tah}. On the other hand, the $m_t\to \infty$ approximation is insufficient for the phenomenological applications. Thereby, many theoretical efforts have been devoted to investigate the finite $m_t$ corrections based on this approximation~\cite{Grigo:2013rya,Frederix:2014hta,Maltoni:2014eza,Grigo:2015dia,Degrassi:2016vss,Grazzini:2018bsd,Davies:2019xzc}.

\vspace{-0.5cm}
\section{N$^3$LO QCD corrections in the infinite top quark mass limit}
\vspace{-0.3cm}

Before discussing our N$^3$LO results, we need to briefly review a few technical aspects of the calculation. The three-loop and four-loop Wilson coefficients for the effective interactions between one and two Higgs bosons and gluons from top-quark loops can be found in Refs.~\cite{Chetyrkin:2005ia, Schroder:2005hy,Kniehl:2006bg,deFlorian:2013uza, Grigo:2014jma, Baikov:2016tgj,Spira:2016zna,Gerlach:2018hen}. According to the number of effective vertices after squaring amplitudes, $gg\to hh$ can be organised into three classes: (a) two-, (b) three-, (c) four-effective vertices. The class (a) shares the same topology as $gg\to h^{(*)}$, and therefore we can recycle the existing N$^3$LO calculation of the single Higgs process~\cite{Anastasiou:2015vya,Anastasiou:2016cez,Mistlberger:2018etf,Dulat:2018rbf}. Due to the power counting of $\alpha_s$, we need to carry out NNLO corrections to the class (b), where the two-loop amplitude has been computed in Ref.~\cite{Banerjee:2018lfq}. We adopt the NNLO $q_T$ subtraction/slicing approach~\cite{Catani:2007vq} to deal with infrared divergences, where some universal ingredients are from Refs.~\cite{Gehrmann:2012ze, Gehrmann:2014yya,Luebbert:2016itl,Echevarria:2016scs, Li:2016ctv,Luo:2019bmw}, while the NLO corrections to di-Higgs plus a jet, which also constitute parts of the computation, were evaluated with the help of the {\sc\small MadGraph5\_aMC@NLO} framework~\cite{Alwall:2014hca,Frederix:2018nkq}. Because the lowest order of the class (c) is $\mathcal{O}(\alpha_s^4)$, we only need to compute the NLO QCD corrections, which were taken care of by {\sc\small MadGraph5\_aMC@NLO} too. The validation of our results is guaranteed by several careful cross checks we have done as shown in Ref.~\cite{Chen:2019fhs}.

\begin{figure}[!ht]
\vspace{-1.0cm}
  \centering
  \subfloat[Cross sections with $m_t=\infty$ limit]{\includegraphics[width=0.30\textwidth,draft=false]{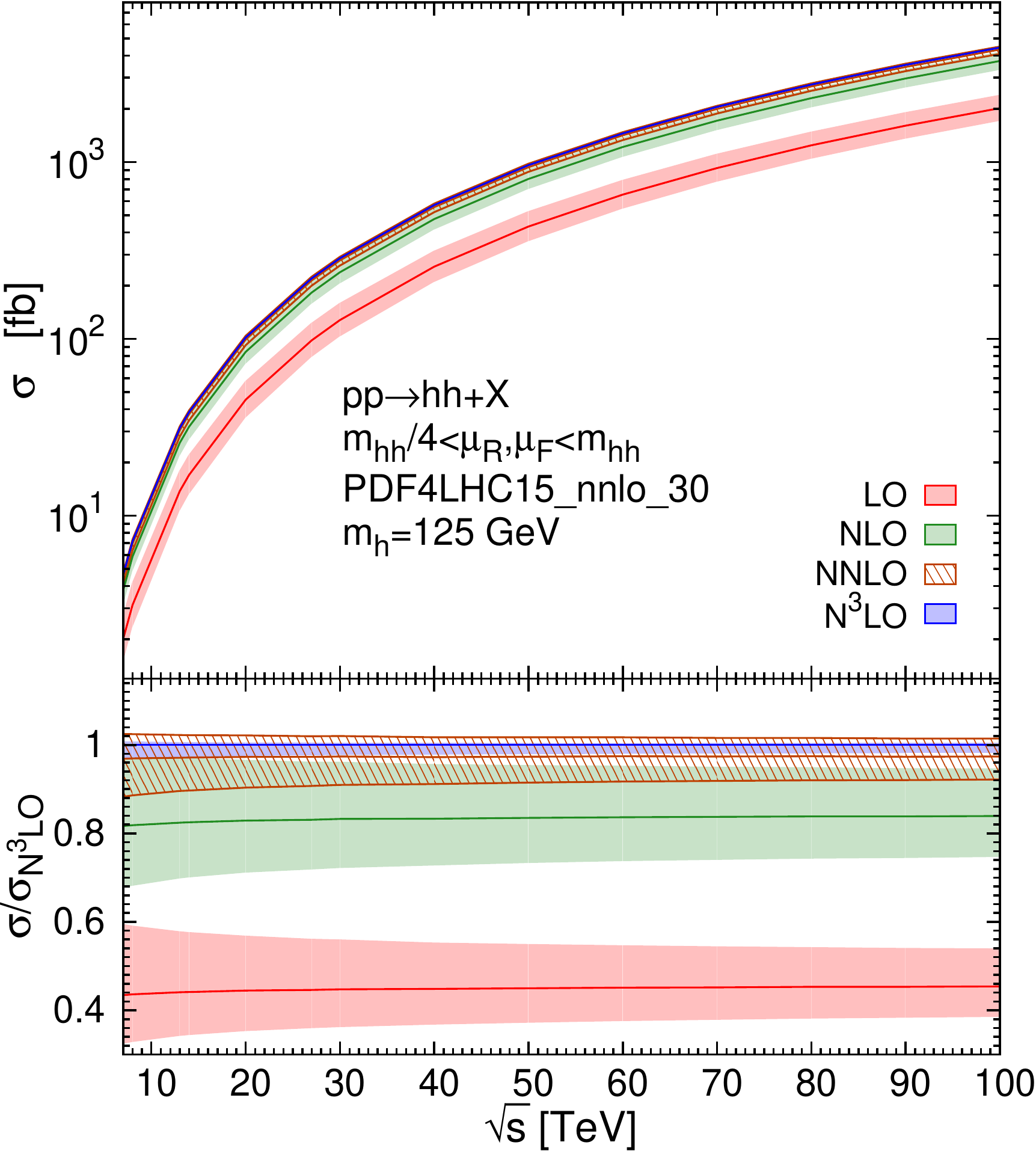}\label{fig:HHXS}}
  \subfloat[Three finite $m_t$ approximations]{\includegraphics[width=0.30\textwidth,draft=false]{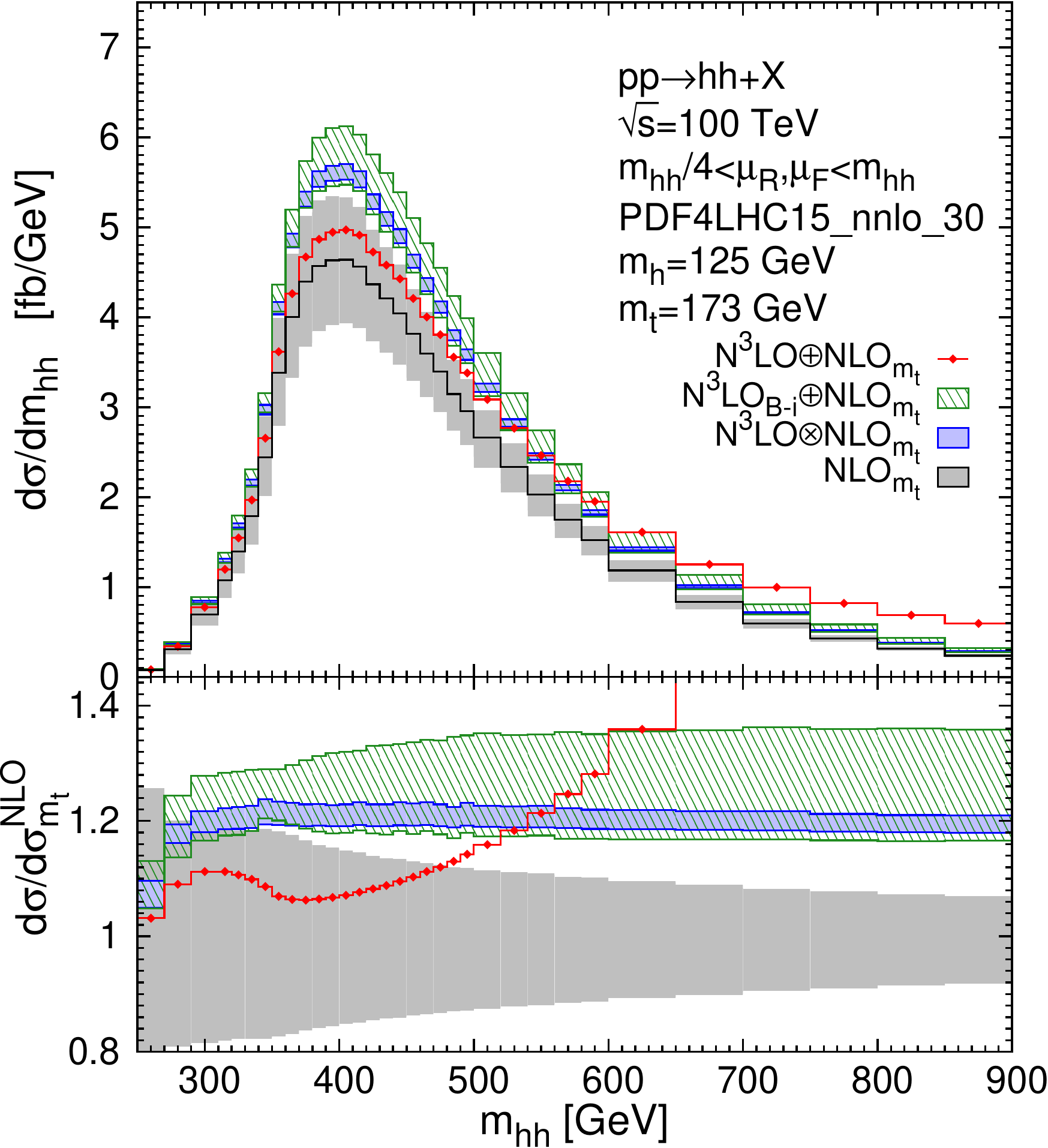}\label{fig:topscheme}}
  \subfloat[NNLO vs N$^3$LO with finite $m_t$]{\includegraphics[width=0.30\textwidth,draft=false]{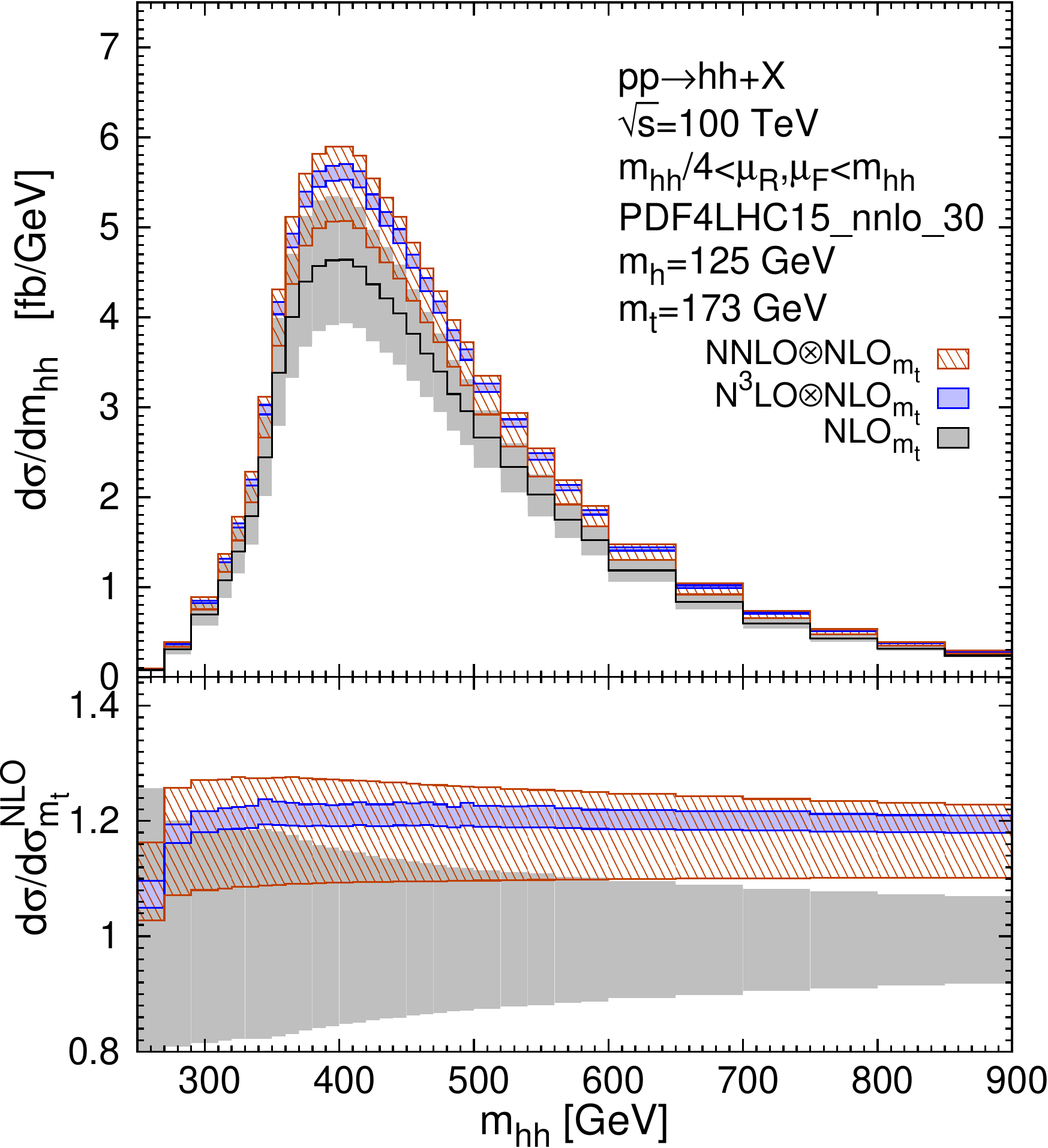}\label{fig:nnlovsn3lo}}
  \vspace{-0.3cm}
  \caption{(a) Total cross sections vs $\sqrt{s}$ from LO to N$^3$LO in the $m_t=\infty$ limit. (b) The comparisons of three $m_t$ approximations at N$^3$LO. (c) NNLO vs N$^3$LO with a $m_t$ approximation. Plots from Refs.~\cite{Chen:2019lzz,Chen:2019fhs}.
  \label{fig:HHatN3LO}}
  \vspace{-0.4cm}
\end{figure}

Figure~\ref{fig:HHXS} reports the total cross sections of $gg\to hh$ in the infinite top quark mass limit as a function of the centre-of-mass energy $\sqrt{s}$ from the LHC energies to $100$ TeV. Results with LO (red), NLO (green), NNLO (brown hatched), and N$^3$LO (blue) accuracies are shown. In the lower inset, the ratios of these cross sections over the central N$^3$LO cross section are displayed. The conventional renormalisation and factorisation scale uncertainties  are represented by the bands. It turns out that the N$^3$LO QCD corrections enhance the NNLO cross sections by around 3\%, while the scale uncertainty is significantly reduced by almost a factor of four. The remaining N$^3$LO scale uncertainty is below the parton distribution function (PDF) parameterisation uncertainty. Meanwhile, we observe very good (asymptotic) perturbative convergence at this order. Besides the total cross sections, the Higgs pair invariant mass distributions are also known exactly at N$^3$LO. Other differential distributions are only presented in an approximated way. This is because the N$^3$LO corrections to the class (a) are not available, and we have to use the global N$^3$LO over NNLO K factors for the partial results of this class (a). We refer the interested readers to Ref.~\cite{Chen:2019fhs} for more details. The overall message is that for all the differential distributions we have considered, except a few regimes that populated by soft-gluon radiation, N$^3$LO changes their shapes in a very mild way. This gives us the confidence of the asymptotic $\alpha_s$ perturbative convergence in the process.

\vspace{-0.5cm}
\section{Finite top quark mass corrections at N$^3$LO}
\vspace{-0.3cm}

The inclusion of finite top quark mass corrections are essential to scrutinise the (differential) cross sections of $gg\to hh$. Since the full $m_t$ results have been computed only at NLO, we have to take into account the finite $m_t$ corrections in the NNLO and N$^3$LO corrections. We have used three different top quark mass approximation schemes~\cite{Chen:2019fhs}. A comparison of them can be found in Figure~\ref{fig:topscheme}, where the full $m_t$ NLO result (NLO$_{m_t}$,grey) from {\sc\small Powheg-Box}~\cite{Heinrich:2017kxx,Heinrich:2019bkc,Alioli:2010xd} is also shown. Our best prediction is that obtained within the N$^3$LO$\otimes$NLO$_{m_t}$ scheme (blue band), where the N$^3$LO in the $m_t=\infty$ limit has been rescaled with full $m_t$ NLO. One can attribute to the difference between N$^3$LO$\otimes$NLO$_{m_t}$ and N$^3$LO$_{{\rm B-i}} \oplus$NLO (green hatched) as a way to estimate the missing top quark mass uncertainty. Such an uncertainty is around 5\% at N$^3$LO. Figure~\ref{fig:nnlovsn3lo} compares NNLO and N$^3$LO results with our best finite $m_t$ scheme. As it should be, this scheme inherits the scale uncertainties of the infinite  $m_t$ results. Therefore, as expected, the band width of N$^3$LO$\otimes$NLO$_{m_t}$ (blue) is significantly reduced with respect to that of NNLO$\otimes$NLO$_{m_t}$ (brown hatched). Again, we refer to the interested readers to Ref.~\cite{Chen:2019fhs} for the complete results, i.e., at other centre-of-mass energies and/or for other differential distributions.


\vspace{-0.5cm}
\section{Conclusions}
\vspace{-0.3cm}

In this talk, we have reported the first N$^3$LO calculations for Higgs pair production in the gluon fusion channel with the infinite top quark mass limit. At this order, the scale uncertainty has been significantly reduced to be below 3\% (2\%) at $13$ ($100$) TeV, where the PDF parameterisation uncertainty is bigger than the scale uncertainty. Our results show pretty good asymptotic $\alpha_s$ perturbative convergence at N$^3$LO. For the purpose of phenomenological applications, we have also taken into account the approximated finite top quark mass effects at NNLO and N$^3$LO together with the known full $m_t$ NLO results. It provides the state-of-the-art predictions for the future di-Higgs measurements at the LHC and at other proposed high-energy accelerators.

{\bf \emph{Acknowledgements.}}
HSS is supported by the European Union's Horizon 2020 research and innovation programme under grant agreement No. 824093 to the EU Virtual Access {\sc\small NLOAccess}.

\bibliographystyle{Science}
\bibliography{reference}

\begin{thebibliography}{10}

\bibitem{Kajantie:1996mn}
K.~Kajantie, M.~Laine, K.~Rummukainen, M.~E. Shaposhnikov, {\it Phys. Rev.
  Lett.\/} {\bf 77}, 2887 (1996).

\bibitem{Csikor:1998eu}
F.~Csikor, Z.~Fodor, J.~Heitger, {\it Phys. Rev. Lett.\/} {\bf 82}, 21 (1999).

\bibitem{Shaposhnikov:1987tw}
M.~Shaposhnikov, {\it Nucl. Phys. B\/} {\bf 287}, 757 (1987).

\bibitem{Witten:1984rs}
E.~Witten, {\it Phys. Rev. D\/} {\bf 30}, 272 (1984).

\bibitem{Hogan:1986qda}
C.~Hogan, {\it Mon. Not. Roy. Astron. Soc.\/} {\bf 218}, 629 (1986).

\bibitem{Grasso:2000wj}
D.~Grasso, H.~R. Rubinstein, {\it Phys. Rept.\/} {\bf 348}, 163 (2001).

\bibitem{Degrassi:2012ry}
G.~Degrassi, {\it et~al.\/}, {\it JHEP\/} {\bf 08}, 098 (2012).

\bibitem{Chang:2019vez}
S.~Chang, M.~A. Luty, {\it JHEP\/} {\bf 03}, 140 (2020).

\bibitem{deBlas:2019rxi}
J.~de~Blas, {\it et~al.\/}, {\it JHEP\/} {\bf 01}, 139 (2020).

\bibitem{Mangano:2020sao}
M.~L. Mangano, G.~Ortona, M.~Selvaggi  (2020).

\bibitem{Borowka:2016ehy}
S.~Borowka, {\it et~al.\/}, {\it Phys. Rev. Lett.\/} {\bf 117}, 012001 (2016).
  [Erratum: Phys. Rev. Lett.117,no.7,079901(2016)].

\bibitem{Borowka:2016ypz}
S.~Borowka, {\it et~al.\/}, {\it JHEP\/} {\bf 10}, 107 (2016).

\bibitem{Baglio:2018lrj}
J.~Baglio, {\it et~al.\/}, {\it Eur. Phys. J.\/} {\bf C79}, 459 (2019).

\bibitem{Davies:2019dfy}
J.~Davies, {\it et~al.\/}, {\it JHEP\/} {\bf 11}, 024 (2019).

\bibitem{Heinrich:2017kxx}
G.~Heinrich, S.~P. Jones, M.~Kerner, G.~Luisoni, E.~Vryonidou, {\it JHEP\/}
  {\bf 08}, 088 (2017).

\bibitem{Jones:2017giv}
S.~Jones, S.~Kuttimalai, {\it JHEP\/} {\bf 02}, 176 (2018).

\bibitem{Heinrich:2019bkc}
G.~Heinrich, S.~P. Jones, M.~Kerner, G.~Luisoni, L.~Scyboz, {\it JHEP\/} {\bf
  06}, 066 (2019).

\bibitem{Baglio:2020ini}
J.~Baglio, {\it et~al.\/}, {\it JHEP\/} {\bf 04}, 181 (2020).

\bibitem{Maltoni:2014eza}
F.~Maltoni, E.~Vryonidou, M.~Zaro, {\it JHEP\/} {\bf 11}, 079 (2014).

\bibitem{Dawson:1998py}
S.~Dawson, S.~Dittmaier, M.~Spira, {\it Phys. Rev.\/} {\bf D58}, 115012 (1998).

\bibitem{deFlorian:2013uza}
D.~de~Florian, J.~Mazzitelli, {\it Phys. Lett.\/} {\bf B724}, 306 (2013).

\bibitem{deFlorian:2013jea}
D.~de~Florian, J.~Mazzitelli, {\it Phys. Rev. Lett.\/} {\bf 111}, 201801
  (2013).

\bibitem{Grigo:2014jma}
J.~Grigo, K.~Melnikov, M.~Steinhauser, {\it Nucl. Phys.\/} {\bf B888}, 17
  (2014).

\bibitem{deFlorian:2016uhr}
D.~de~Florian, {\it et~al.\/}, {\it JHEP\/} {\bf 09}, 151 (2016).

\bibitem{Chen:2019lzz}
L.-B. Chen, H.~T. Li, H.-S. Shao, J.~Wang, {\it Phys. Lett.\/} {\bf B803},
  135292 (2020).

\bibitem{Chen:2019fhs}
L.-B. Chen, H.~T. Li, H.-S. Shao, J.~Wang, {\it JHEP\/} {\bf 03}, 072 (2020).

\bibitem{Shao:2013bz}
D.~Y. Shao, C.~S. Li, H.~T. Li, J.~Wang, {\it JHEP\/} {\bf 07}, 169 (2013).

\bibitem{deFlorian:2015moa}
D.~de~Florian, J.~Mazzitelli, {\it JHEP\/} {\bf 09}, 053 (2015).

\bibitem{deFlorian:2018tah}
D.~De~Florian, J.~Mazzitelli, {\it JHEP\/} {\bf 08}, 156 (2018).

\bibitem{Grigo:2013rya}
J.~Grigo, J.~Hoff, K.~Melnikov, M.~Steinhauser, {\it Nucl. Phys.\/} {\bf B875},
  1 (2013).

\bibitem{Frederix:2014hta}
R.~Frederix, {\it et~al.\/}, {\it Phys. Lett.\/} {\bf B732}, 142 (2014).

\bibitem{Grigo:2015dia}
J.~Grigo, J.~Hoff, M.~Steinhauser, {\it Nucl. Phys.\/} {\bf B900}, 412 (2015).

\bibitem{Degrassi:2016vss}
G.~Degrassi, P.~P. Giardino, R.~Grober, {\it Eur. Phys. J.\/} {\bf C76}, 411
  (2016).

\bibitem{Grazzini:2018bsd}
M.~Grazzini, {\it et~al.\/}, {\it JHEP\/} {\bf 05}, 059 (2018).

\bibitem{Davies:2019xzc}
J.~Davies, F.~Herren, G.~Mishima, M.~Steinhauser, {\it JHEP\/} {\bf 05}, 157
  (2019).

\bibitem{Chetyrkin:2005ia}
K.~G. Chetyrkin, J.~H. Kuhn, C.~Sturm, {\it Nucl. Phys.\/} {\bf B744}, 121
  (2006).

\bibitem{Schroder:2005hy}
Y.~Schroder, M.~Steinhauser, {\it JHEP\/} {\bf 01}, 051 (2006).

\bibitem{Kniehl:2006bg}
B.~A. Kniehl, A.~V. Kotikov, A.~I. Onishchenko, O.~L. Veretin, {\it Phys. Rev.
  Lett.\/} {\bf 97}, 042001 (2006).

\bibitem{Baikov:2016tgj}
P.~A. Baikov, K.~G. Chetyrkin, J.~H. Kuhn, {\it Phys. Rev. Lett.\/} {\bf 118},
  082002 (2017).

\bibitem{Spira:2016zna}
M.~Spira, {\it JHEP\/} {\bf 10}, 026 (2016).

\bibitem{Gerlach:2018hen}
M.~Gerlach, F.~Herren, M.~Steinhauser, {\it JHEP\/} {\bf 11}, 141 (2018).

\bibitem{Anastasiou:2015vya}
C.~Anastasiou, C.~Duhr, F.~Dulat, F.~Herzog, B.~Mistlberger, {\it Phys. Rev.
  Lett.\/} {\bf 114}, 212001 (2015).

\bibitem{Anastasiou:2016cez}
C.~Anastasiou, {\it et~al.\/}, {\it JHEP\/} {\bf 05}, 058 (2016).

\bibitem{Mistlberger:2018etf}
B.~Mistlberger, {\it JHEP\/} {\bf 05}, 028 (2018).

\bibitem{Dulat:2018rbf}
F.~Dulat, A.~Lazopoulos, B.~Mistlberger, {\it Comput. Phys. Commun.\/} {\bf
  233}, 243 (2018).

\bibitem{Banerjee:2018lfq}
P.~Banerjee, S.~Borowka, P.~K. Dhani, T.~Gehrmann, V.~Ravindran, {\it JHEP\/}
  {\bf 11}, 130 (2018).

\bibitem{Catani:2007vq}
S.~Catani, M.~Grazzini, {\it Phys. Rev. Lett.\/} {\bf 98}, 222002 (2007).

\bibitem{Gehrmann:2012ze}
T.~Gehrmann, T.~Lubbert, L.~L. Yang, {\it Phys. Rev. Lett.\/} {\bf 109}, 242003
  (2012).

\bibitem{Gehrmann:2014yya}
T.~Gehrmann, T.~Luebbert, L.~L. Yang, {\it JHEP\/} {\bf 06}, 155 (2014).

\bibitem{Luebbert:2016itl}
T.~Lubbert, J.~Oredsson, M.~Stahlhofen, {\it JHEP\/} {\bf 03}, 168 (2016).

\bibitem{Echevarria:2016scs}
M.~G. Echevarria, I.~Scimemi, A.~Vladimirov, {\it JHEP\/} {\bf 09}, 004 (2016).

\bibitem{Li:2016ctv}
Y.~Li, H.~X. Zhu, {\it Phys. Rev. Lett.\/} {\bf 118}, 022004 (2017).

\bibitem{Luo:2019bmw}
M.-X. Luo, T.-Z. Yang, H.~X. Zhu, Y.~J. Zhu, {\it JHEP\/} {\bf 01}, 040 (2020).

\bibitem{Alwall:2014hca}
J.~Alwall, {\it et~al.\/}, {\it JHEP\/} {\bf 07}, 079 (2014).

\bibitem{Frederix:2018nkq}
R.~Frederix, {\it et~al.\/}, {\it JHEP\/} {\bf 07}, 185 (2018).

\bibitem{Alioli:2010xd}
S.~Alioli, P.~Nason, C.~Oleari, E.~Re, {\it JHEP\/} {\bf 06}, 043 (2010).

\end{thebibliography}



\end{document}